\documentclass[prd,jmp,reprint,onecolumn,superscriptaddress,12pt]{revtex4-2}

\usepackage{amssymb}
\usepackage{amsmath}
\usepackage{graphicx}
\usepackage{dcolumn}
\usepackage{bm}
\usepackage{color}
\usepackage{xcolor}
\usepackage{float}
\allowdisplaybreaks
\usepackage{subfig}
\usepackage{comment}
\usepackage{multirow}
\usepackage[figurename=FIGURE,tablename=TABLE]{caption}
\usepackage{threeparttable}
\usepackage{hyperref}
\hypersetup{
    colorlinks=true, 
    linktoc=all,     
    linkcolor=magenta,
    citecolor=red
}

\newcommand{\ba}{\begin{eqnarray}}
\newcommand{\ea}{\end{eqnarray}}

\begin{document}

\title{Shadows and photon spheres in static and rotating traversable wormholes}

\author{Takol Tangphati} 
\email{takol.ta@mail.wu.ac.th}
\affiliation{School of Science, Walailak University, Nakhon Si Thammarat, 80160, Thailand}

\author{Phongpichit Channuie}
\email{phongpichit.ch@mail.wu.ac.th}
\affiliation{School of Science, Walailak University, Nakhon Si Thammarat, 80160, Thailand}
\affiliation{College of Graduate Studies, Walailak University, Nakhon Si Thammarat, 80160, Thailand}

\author{Kazuharu Bamba}
\email{bamba@sss.fukushima-u.ac.jp}
\affiliation{Faculty of Symbiotic Systems Science, Fukushima University, Fukushima 960-1296, Japan}

\author{Davood Momeni}
\email{dmomeni@nvcc.edu}
\affiliation{Northern Virginia Community College, 8333 Little River Turnpike, Annandale, VA 22003, USA}

\date{\today}

\begin{abstract}

The shadows and photon spheres are investigated for both static and rotating traversable wormholes. In particular, with a red-shift function constructed in combination with two distinct shape functions, the null geodesic equation is found to predict the light trajectory. To explore observable phenomena and gain a deeper understanding of the gravitational influences originating from wormholes, we explicitly demonstrate the shadow produced and photon spheres associated with both static and rotating traversable wormholes by taking into account the gravitational lensing effects and photon motion in strong gravitational fields. Specifically, the ray tracing of the light geodesic and profiles of the intensity around both static and rotating traversable wormholes can be accomplished. 

\end{abstract}


\maketitle

\newpage
\section{Introduction}

Wormholes, hypothetical structures in spacetime, have long been a subject of fascination and exploration in the realm of theoretical physics and astrophysics. These traversable tunnels through spacetime, while primarily a theoretical construct, continue to captivate the scientific community due to their potential role in understanding the fundamental laws of the universe, particularly in the context of Einstein's theory of general relativity. The concept of wormholes had its theoretical foundations initially established by Flamm in 1916~\cite{Flamm:1916}. Subsequently, a detailed study of the wormholes has been further extended by Albert Einstein and Nathan Rosen in their pioneering paper in 1935 \cite{Einstein:1935tc}, often referred to as the ``Einstein-Rosen bridge." The study of black holes and wormholes converged in the 20th century as physicists explored deeper into the mathematical and physical underpinnings of these exotic objects.

One of the most widely recognized papers that contributed to our understanding of wormholes was published in 1988. This paper explored the theoretical aspects of traversable wormholes and their potential applications in interstellar travel \cite{Morris:1988cz}. This work not only brought wormholes into the mainstream of astrophysical research but also introduced the concept of "exotic matter" as a key ingredient for stabilizing these hypothetical passages through spacetime. In the literature, many authors have intensively studied various aspects of traversable wormhole (TW) geometries, e.g., Refs.~\cite{Penington:2019kki,Almheiri:2019qdq,Hawking:1988ae,Visser:1989kh,Lemos:2003jb,Bronnikov:2002rn}. More specifically, a natural scenario for the existence of the exotic energies can sustain traversable wormholes. For example, phantom energy~\cite{Lobo:2005us,Sushkov:2005kj}. Moreover, there have been some other interesting wormhole scenarios appeared in the literature, e.g., teleparralel gravity~\cite{Boehmer:2012uyw,Mustafa:2021ykn} or $f(T)$ gravity~\cite{Sharif:2013exa,Jamil:2012ti} (the recent extesion of which is called as $f(Q)$ gravity~\cite{Heisenberg:2023lru}), $f(R)$ gravity~\cite{Lobo:2009ip,Furey:2004rq} (see Refs.~\cite{Sotiriou:2008rp,DeFelice:2010aj,Nojiri:2010wj,Capozziello:2011et,Clifton:2011jh,Joyce:2014kja,Nojiri:2017ncd} for models of $f(R)$ and modified gravity), Einstein-Gauss-Bonnet theory~\cite{Bhawal:1992sz,Canate:2022dzb,Kanti:2011jz,Kanti:2011yv,Mehdizadeh:2015jra,Antoniou:2019awm}, bumblebee gravity~\cite{Ovgun:2018xys}, and other interesting scenarios Refs.~\cite{Botta-Cantcheff:2009ffi,Lobo:2005vc,Izmailov:2010pk,Capozziello:2012hr,Biswas:2023ofz,Biswas:2022wah,DeFalco:2020afv,DeFalco:2021klh,DeFalco:2021ksd,DeFalco:2021btn,DeFalco:2023kqy,Elizalde:2018frj,Antonini:2022ptt,Chatzifotis:2021hpg,Banerjee:2021mqk, Nojiri:2023dvf}. Recently, Casimir energy has been considered as a potential source to generate a traversable wormhole~\cite{Garattini:2019ivd}. It is used to proof the existence of negative energy which can be built in the laboratory. Its extension has recently been investigated by many authors, e.g., Refs.~\cite{Jusufi:2020rpw,Garattini:2021kca,Hassan:2022hcb,Samart:2021tvl,Garattini:2023qyo}. Perhaps, dark energy could also be used as a source to stabilize the wormholes, see e.g., Refs.~\cite{Nojiri:2006ri,Copeland:2006wr,Durrer:2007re,Bamba:2012cp}.

Among the many intriguing aspects of wormholes, the study of their shadows and photon spheres offers a unique lens through which we can investigate their properties and implications. The concept of a ``shadow" in astrophysics refers to the apparent shape of an object as seen from a distant observer. This phenomenon is closely related to gravitational lensing, a consequence of the bending of light around massive objects. While shadows have been extensively studied in the context of black holes, see, e.g., Ref.~\cite{Gralla:2019xty} and references therein, the investigation of shadows in wormholes is an emerging field with distinct challenges and opportunities. Shadows cast by wormholes can reveal information about their size, shape, and the distribution of exotic matter within them, shedding light on the feasibility of traversable wormholes and their potential observational signatures.

Photon spheres, on the other hand, are regions in the vicinity of massive objects where photons can orbit in stable, closed paths. The existence and properties of photon spheres are intimately connected to the spacetime geometry surrounding these objects. The study of photon spheres in wormholes can provide insights into the gravitational and geometric characteristics of these structures. Understanding the differences between rotating and non-rotating wormholes in terms of their photon spheres is essential, as rotation can profoundly affect the spacetime curvature and, consequently, the behavior of light within their vicinity.

It has been found that the shadow of rotating traversable wormholes has been stuided in Refs.~\cite{Nedkova:2013msa,Gyulchev:2018fmd}. More specifically, in Ref.~\cite{Nedkova:2013msa}, the authors considered a couple of wormhole solutions and examine the structure of their shadow images, and the intrinsic mechanisms for their formation; while, Ref.~\cite{Gyulchev:2018fmd} took the same wormhole solutions and studied the images. Having compared to the present work, we will instead consider different red-shift functions and study shadows and photon sphere taking into account the profile intensity observed by the observer as well as the ray tracing pattern.

In the present paper, we analyze shadows and photon spheres in both static and rotating traversable wormholes. By employing mathematical formalism and numerical simulations, we investigate the gravitational lensing effects and photon motion in strong gravitational fields. 

The organization of this paper is as follows. In Sec.~\ref{ch1}, we explain the mathematical formalism of the null geodesic equation allowing to predict the light trajectory obtained by the Euler-Lagrange equation. We also formulate a specific intensity observed by the observer. In Sec.~\ref{Num}, we use numerical techniques to explicitly depict the shadow produced and photon spheres associated with both static and rotating traversable wormholes. Finally, we present discussions and conclusions including more insights and further outlooks in Sec.~\ref{conc}.

\section{Light deflection in Traversable Wormholes}\label{ch1}
Traversable wormholes, like other massive objects in space, can cause gravitational lensing. This means their gravitational fields can bend the path of light as it passes near the wormhole. Understanding the deflection of light by a wormhole's gravitational field is crucial for astronomers and astrophysicists. Studying light deflection in the context of traversable wormholes contributes to our understanding of the fundamental principles of general relativity and the behavior of light in extreme gravitational environments. 

\subsection{Static Case}

The idea of a traversable wormhole was initially proposed by Morris and Thorne in 1988 \cite{Morris:1988cz}. In contrast to earlier concepts such as the Einstein-Rosen bridge or Wheeler's microscopic charge-carrying wormholes, which either allowed only one-way travel or lacked the capability for objects like humans to pass through, traversable wormholes explicitly enable two-way travel. We start by considering a static, spherically symmetric and asymptotically flat space-time with the metric
\begin{eqnarray}\label{met}
    ds^2_{\text{stat}} = -e^{2 \Phi} dt^2 + \frac{dr^2}{1- \frac{b}{r}} + r^2 (d \theta^2 + \sin^2 \theta d \phi^2),
\end{eqnarray}
where $\Phi$ and $b$ are arbitrary functions of the variable `$r$,' where $\Phi$ is referred to as the `redshift function', and `$b$' is known as the `shape function. The redshift function $\Phi$ is responsible for determining how the gravitational redshift behaves as an object falls into the gravitational field, while the shape function `$b$' describes the geometrical characteristics of the wormhole when visualized on an embedding diagram. A previous study, as cited in Ref.~\cite{Morris:1988cz}, has demonstrated that for the metric to represent a valid wormhole, the function `$b$' must adhere to a specific condition known as the `flare-out' condition. When this condition is met, Eq.~(\ref{met}) describes a scenario where two identical asymptotic universes are connected at a point known as the 'throat,' located at $r=b$.' Ensuring the wormhole's traversability is critical and implies the absence of event horizons or curvature singularities. To meet this requirement, it is necessary for the function `$\Phi$' to remain finite at all points within the system. To investigate the light deflection influenced by the wormholes, we must quantify how the light ray moves around them. Basically, we consider the path of a light ray following the null geodesic equation, allowing us to predict the light trajectory. The equation can be obtained considering the Euler-Lagrange equation: \cite{Zeng:2020dco}
\begin{eqnarray}
    \frac{d}{d \lambda} \left( \frac{\partial \mathcal{L}}{\partial \dot{x}^{\mu}} \right) = \frac{\partial \mathcal{L}}{\partial x^{\mu}},
\end{eqnarray}
where 
\begin{eqnarray}
    \mathcal{L} = \frac{1}{2} g_{\mu \nu} \dot{x}^{\mu} \dot{x}^{\nu} = \frac{1}{2} \left( -e^{2\Phi} \dot{t}^2 + \frac{\dot{r}^2}{1 - \frac{b}{r}} + r^2 \dot{\theta}^2 + r^2 \sin^2 \theta \dot{\phi}^2 \right),
    \label{lagrangian_staic}
\end{eqnarray}
and the $\, \dot{} \, \equiv \frac{d}{d \lambda}$ represents the derivative over an affine parameter $\lambda$.

In our case, we consider the equatorial plane with $\theta = \pi/2$ and apply the variational principle method to the Lagrangian for the static traversable wormhole in Eq.~(\ref{lagrangian_staic}). We obtain 
\begin{eqnarray}
    \dot{t} &=& \frac{E}{e^{2 \Phi}}, \label{time_static} \\
    \dot{\phi} &=& \frac{L}{r^2}, \label{phi_static}
\end{eqnarray}
where $E$ and $L$ is the total energy and the angular momentum of light, respectively. We consider the null geodesic $(\mathcal{L} = 0)$ which can be rewritten in term of the kinetic and potential energy and re-scale an affine parameter $\lambda \rightarrow \lambda / |L|$
\begin{eqnarray}
    T + V = \frac{1}{\Tilde{b}^2}, \label{energy_conserv_static}
\end{eqnarray}
where we have defined
\begin{eqnarray}
    T &\equiv& \frac{e^{2 \Phi}}{1 - \frac{b}{r}} \dot{r}^2, \\
    V &\equiv& \frac{e^{2 \Phi}}{r^2},
\end{eqnarray}
and $\Tilde{b} \equiv \frac{L}{|E|}$ is the impact parameter. 
To find the photon sphere, we apply the relations $\dot{r} = 0$ and $\ddot{r} = 0$ on Eq.~(\ref{energy_conserv_static}) and obtain the photon sphere radius, $r_{\text{ph}}$, from the following equation
\begin{eqnarray}
    e^{2\Phi} \Tilde{b}^2 \bigg\vert_{r_{\text{ph}}} = r_{\text{ph}}^2
\end{eqnarray}

The trajectory of the null geodesic can be expressed by the combination of the equations of motion in Eq.~(\ref{phi_static}) and Eq.~(\ref{energy_conserv_static}) as shown
\begin{eqnarray}
    \left( \frac{dr}{d \phi} \right)^2 = \left( 1 - \frac{b}{r} \right) \left(  \frac{r^4}{\Tilde{b}^2 e^{2\Phi}} - r^2 \right). \label{null_static}
\end{eqnarray}
With the numerical technique to solve the null geodesic in Eq.~(\ref{null_static}), we substitute the reciprocal $u \equiv 1/r$ and rewrite Eq.~(\ref{null_static}) as
\begin{eqnarray}\label{u_phi_F}
    \left( \frac{du}{d \phi} \right)^2 =  \left( 1 - b u \right) \left( \frac{1}{\Tilde{b}^2 e^{2\Phi}} - u^2 \right)\equiv {\cal F}(u)\,.
\end{eqnarray}
From Eq.(\ref{u_phi_F}), we can solve $\phi$ with respect to $u$. We can plot the trajectory of the light ray in Sec.\ref{Num}.

\subsection{Rotating Case}

In the rotating case, we consider the metric tensor of the rotating traversable wormhole which is of the distinct advantage of making the physics transparent \cite{Hartle:1967,Hartle:1967he} (see also \cite{Teo:1998dp}):
\begin{eqnarray}
    ds^2_{\text{rot}} = - e^{2\Phi} dt^2 + \frac{dr^2}{1 - \frac{b}{r}} + r^2 K^2 \left( d\theta^2 + \sin^2 \theta \left( d \phi - \omega dt\right)^2 \right),
\end{eqnarray}
where the four functions $\Phi,\,K$ and $\omega$ may ingeneral depend on $(r,\theta)$. However, in this work we assume that $K = K(r)$ is an arbitrary function of $r$ and $\omega = \omega(r)$ is the angular velocity of the rotating traversable wormhole. In this work, we assume $\omega(r)\sim 2 a/r^3$ \cite{Teo:1998dp} with $a$ is the total angular momentum of the wormhole. The Euler-Lagrane equation for this case reads
\begin{eqnarray}
    \mathcal{L} &=& \frac{1}{2} g_{\mu \nu} \dot{x}^{\mu} \dot{x}^{\nu} \nonumber\\&=& \frac{1}{2} \left( - e^{2\Phi} \dot{t}^2 + \frac{\dot{r}^2}{1 - \frac{b}{r}} + r^2 \left( \dot{\theta}^2 + \sin^2 \theta \dot{\phi}^2 \right) + r^2 \sin^2 \theta \left[ - 2 \omega \dot{\phi} \dot{t} + \omega^2 \dot{t}^2 \right] \right)\,. \label{Lagrane_rotating}
\end{eqnarray}

The equations of motion can be straightforwardly derived from the Lagrangian in Eq.~(\ref{Lagrane_rotating}). We find
\begin{eqnarray}
    \left( - e^{2\Phi} + r^2 K^2 \omega^2  \right) \dot{t} - \omega r^2 K^2 \dot{\phi} &=& - E\,. \label{energy_rotating} \\
    r^2 K^2 (\dot{\phi} - \omega \dot{t})&=& L\,. \label{angular_rotating}
\end{eqnarray}
The null geodesic ($\mathcal{L} = 0$) for this case with the re-scale $\lambda \rightarrow \lambda/|L|$ reads
\begin{eqnarray}
    \Tilde{T} + \Tilde{V} = \frac{1}{\Tilde{b}^2}, \label{energy_conservation_rotating}
\end{eqnarray}
where
\begin{eqnarray}
    \Tilde{T} &=& \frac{\dot{r}^2 e^{2 \Phi}}{1 - \frac{b}{r}}, \\
    \Tilde{V} &=& \frac{e^{2 \Phi}}{K^2 r^2} + \frac{2 \omega}{\Tilde{b}}  - \omega^2.
\end{eqnarray}

The photon ring radius for the rotating traversable wormhole, $\Tilde{r}_{\text{ph}}$, satisfies the relation
\begin{eqnarray}
    \Tilde{r}_{\text{ph}} = \frac{\Tilde{b}}{1 - \Tilde{b} \omega} \frac{e^{\Phi}}{K}\bigg\vert_{r_{\text{ph}}}.
\end{eqnarray}

The trajectory of the null geodesic for the rotating traversable wormhole by combining Eq.~(\ref{energy_rotating}) and Eq.~(\ref{angular_rotating}) to obtain
\begin{eqnarray}\label{dpr}
    \left( \frac{dr}{d \phi} \right)^2 = K^4 \left( 1 - \frac{b}{r} \right) \left( \frac{r^4 (1 - b \omega)^2}{e^{2\Phi} b^2} - \frac{r^2}{K^2}\right) \bigg/ \left( 1 + \frac{K^2 r^2 \omega (E -  \omega)}{e^{2 \Phi} } \right)^2\,,
\end{eqnarray}
with $u = 1/r$. The above relation then becomes
\begin{eqnarray}
    \left( \frac{du}{d \phi} \right)^2 = K^4 \left( 1 - bu \right) \left( \frac{(1 - \Tilde{b} \omega)^2}{e^{2\Phi} \Tilde{b}^2} - \frac{u^2}{K^2}\right) \bigg/ \left( 1 + \frac{K^2 \omega (E - \omega)}{u^2 e^{2 \Phi} } \right)^2\equiv{\cal G}(u)\,.\label{metr}
\end{eqnarray}
From Eq.~(\ref{metr}), we can solve $\phi$ with respect to $u$. Notice that the above expression Eq.~(\ref{metr}) reduces to Eq.~(\ref{u_phi_F}) when setting $\omega=0$ and $K=1$.

\section{Shadows and photon spheres: Numerical results}\label{Num}
The observation of a wormhole's shadow can provide indirect evidence for the existence of the wormhole itself. By studying the size and shape of the shadow, scientists can infer properties of the wormhole, such as its mass, spin, and shape; whilst, photon spheres are significant because they are related to the extreme warping of spacetime caused by massive objects. Understanding the existence and properties of photon spheres around a traversable wormhole can help researchers map the gravitational field of the wormhole, which is crucial for and stability. In this work, we consider the redshift function, $\Phi$, and the shape function, $b(r)$, given by
\begin{eqnarray}
    \Phi(r)=\Phi_{0}\log \Big(1-\frac{r_{0}}{r}\Big)^{\alpha}\,, \label{redshift_function}
\end{eqnarray}
and
\begin{eqnarray}
b(r)=\begin{cases}
  {\rm Case}\,\#1: b_{1}(r)=r\exp (-\beta (r-r_{0}))\,, \\
  {\rm Case}\,\#2: b_{2}(r)=r_{0} \left(\frac{r_{0}}{r}\right)^{\gamma}\,,
\end{cases} \label{shape_functions}
\end{eqnarray}
with $\alpha,\,\beta$ and $\gamma$ being arbitrary constants. The shape function $b_{1}(r)$ was suggested in Ref.~\cite{Barros:2018lca} for wormhole geometries supported by three-form fields, whilst $b_{2}(r)$ has been studied in Ref.~\cite{Tangphati:2020mir} for traversable wormholes in $f(R)$-massive gravity. Here we choose $b_{i}(r),\,i=1,2,$ such that $b(r_0) = r_0$ at $r = r_0$. They are also satisfied the flaring-out condition~\cite{Morris:1988cz}, given by $b(r) - b'(r) r \geq 0$, at the vicinity of the throat, where a prime denotes a derivative with respect to the radial coordinate $r$. Additionally, $b(r)/r \rightarrow 0$ as $r \rightarrow \infty$. Note that the additional condition $b(r)/r < 1$ is also satisfied.

\subsection{Static case}
In this section, we formulate the mathematical descriptions of the shadow and photon sphere of the wormholes. Therefore, we have to determine the specific intensity observed by the observer. The observed specific intensity $I$ at the observed photon frequency $I(\nu_0)$ can be ontained by integrating out the specific emissivity along the photon path. The relation reads \cite{Jaroszynski:1997bw,Bambi:2013nla}
\begin{eqnarray}
    I(\nu_0) = \int g^3 j(\nu_e) dl_{\text{prop}},
\end{eqnarray}
where $g \equiv \nu_0 / \nu_e$ is the red shift factor, $\nu_e$ is the photon frequency of the emitter, $\nu_0$ is the observed photon frequency, $dl_{\text{prop}}$ is the infinitesimal proper length, and $j(\nu_e)$ is the emissivity per unit volume measured in the rest frame of the matter. For the static traversable wormholes, we have $g = \nu_0 / \nu_e = e^{\Phi}$. For specific emissivity, we assume \cite{Zeng:2020dco}
\begin{eqnarray}
j(\nu_e) &\propto& \frac{\delta (\nu_e - \nu_r)}{r^2}\,.
\end{eqnarray}
According to Eq.~(\ref{met}), we find the proper length measured in the rest frame of the emitter given by
\begin{eqnarray}
    dl_{\text{prop}} &=& \left( \frac{1}{1 - \frac{b}{r}} + r^2 \left( \frac{d \phi}{dr}\right)^2 \right)^{1/2} dr\,,
\end{eqnarray}
where $d\phi/dr$ is given by Eq.~(\ref{null_static}). However, the intensity with the reciprocal $u(\phi) = 1/r$ can be written as
\begin{eqnarray}
    I(\nu_0) &=& - \int d\phi u'(\phi) e^{3 \Phi} \left( \frac{1}{1 - b u} + \frac{u^2}{u'(\phi)^2} \right)^{1/2}\,.\label{in1}
\end{eqnarray}
We take Eq.~(\ref{in1}) to study the shadow of the static wormholes. It is noticed that the intensity depends on the trajectory of the light ray. Therefore, we can determine a trajectory parameterized by the impact parameter denoted as ``$\tilde{b}$".

\begin{figure}[t]
    \centering
    \includegraphics[width = 7 cm]{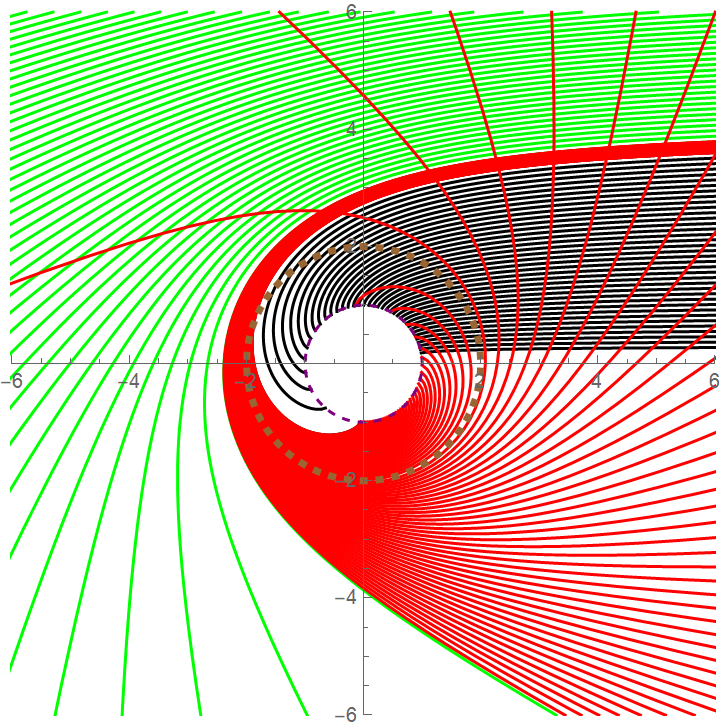}
    \includegraphics[width = 7 cm]{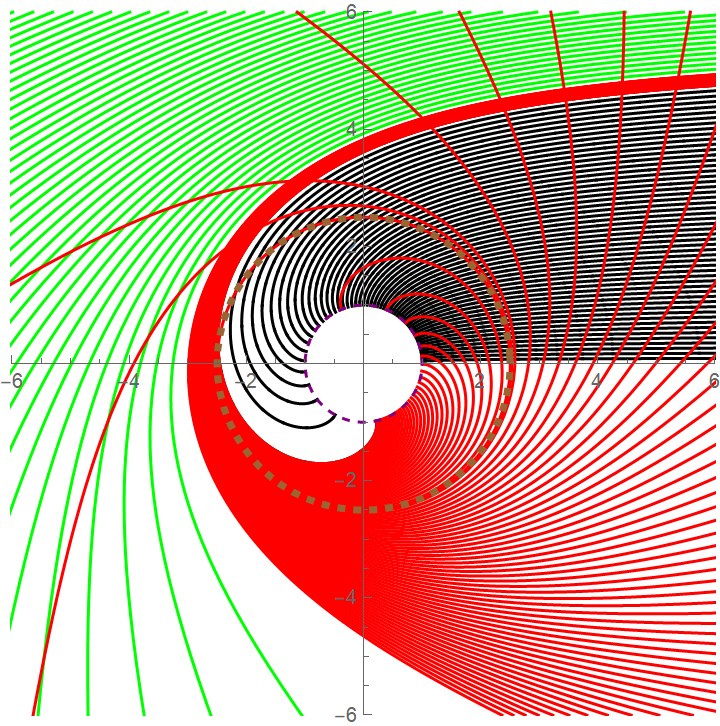}
    \caption{Figures show the ray tracing of the light geodesic around a static traversable wormhole with parameter set: $\alpha = 1, \beta = 1, \Phi_0 = 1$ and $r_0 = 1$, and the shape function $b_{1}(r)$ (Left panel) and $\alpha = 1.5, \beta = 1, \Phi_0 = 1$ and $r_0 = 1$, and the shape function $b_{2}(r)$ (Right panel). The photon ring is symbolized by the dashed circle in brown, while the purple dashed circle indicates the wormhole's throat.}
    \label{fig_static_raytracing1}
\end{figure}

\subsubsection{Ray tracing}
In this case, we illuminate trajectory of the light ray in Fig.\ref{fig_static_raytracing1}. We consider the red-shift function in Eq.~(\ref{redshift_function}) along with the two types of shape function presented in Eq.~(\ref{shape_functions}). For the case of $b_1$, we use a parameter set: $\alpha = 1, \beta = 1, \Phi_0 = 1$ and $r_0 = 1$, and the plot of the ray tracing is displayed in the left panel of Fig.~\ref{fig_static_raytracing1}. For the case of $b_2$, we take a parameter set: $\alpha = 1.5, \gamma = 1, \Phi_0 = 1$ and $r_0 = 1$, and present the ray tracing in the right panel of Fig.~\ref{fig_static_raytracing1}.

The black, red and green lines correspond to $\Tilde{b} < \Tilde{b}_{\rm ph},\, \Tilde{b} = \Tilde{b}_{\rm ph}$ and $\Tilde{b} > \Tilde{b}_{\rm ph}$, respectively. We see for the case of $\Tilde{b} < \Tilde{b}_{\rm ph}$ that the light ray decreases all the way into the wormholes, corresponding to a black region in Fig.~\ref{fig_static_raytracing1}. For the case of $\Tilde{b} > \Tilde{b}_{\rm ph}$, the light ray near the wormholes is reflected back, which corresponds to green region in Fig.~\ref{fig_static_raytracing1}. Notice that for $\Tilde{b} > \Tilde{b}_{\rm ph}$, a turning point emerges, where the light ray changes its radial direction. Such a point is determined by ${\cal F}(u)=0$, where ${\cal F}(u)$ was already given in Eq.~(\ref{u_phi_F}). The photon ring is symbolized by the dashed circle in brown, while the purple dashed circle indicates the wormhole's throat. We found that a critical value of the impact parameter of the case \#1, $\Tilde{b}_{ph, 1}=4.0$, which is less than that of the case \#2, $\Tilde{b}_{ph, 2}=5.3$.

\subsubsection{Intensity}

\begin{figure}[h]
    \centering
    \includegraphics[width = 7 cm]{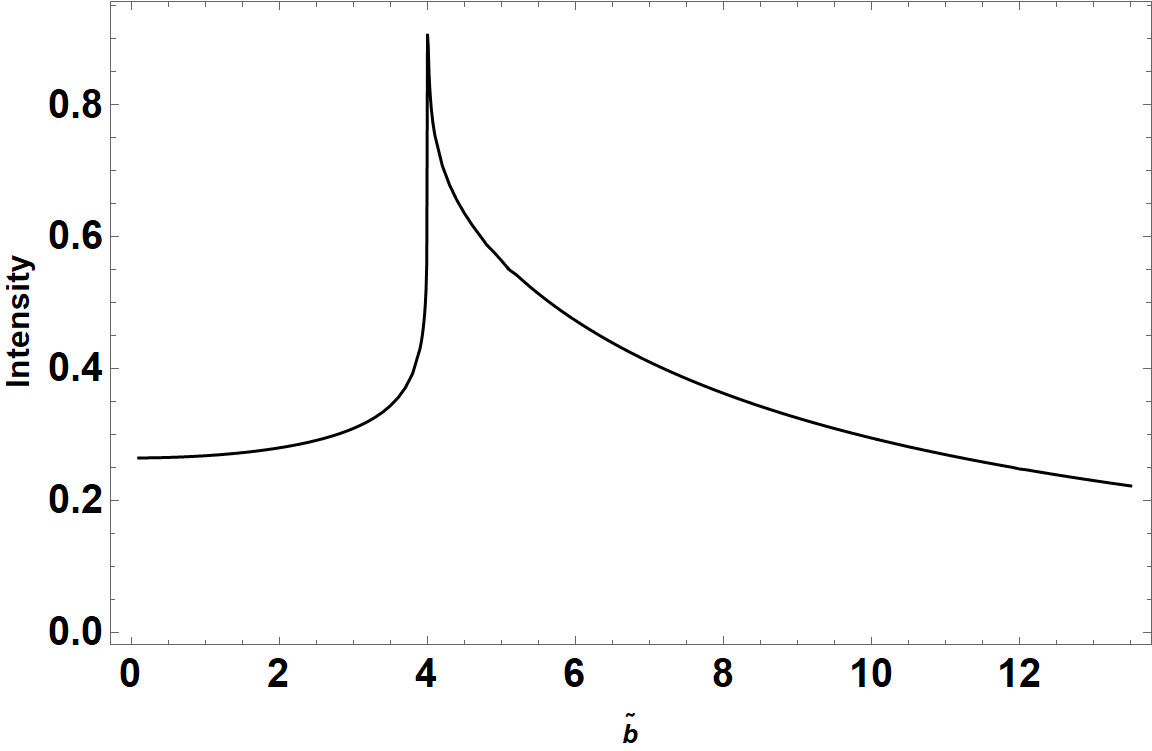}
    \includegraphics[width = 7 cm]{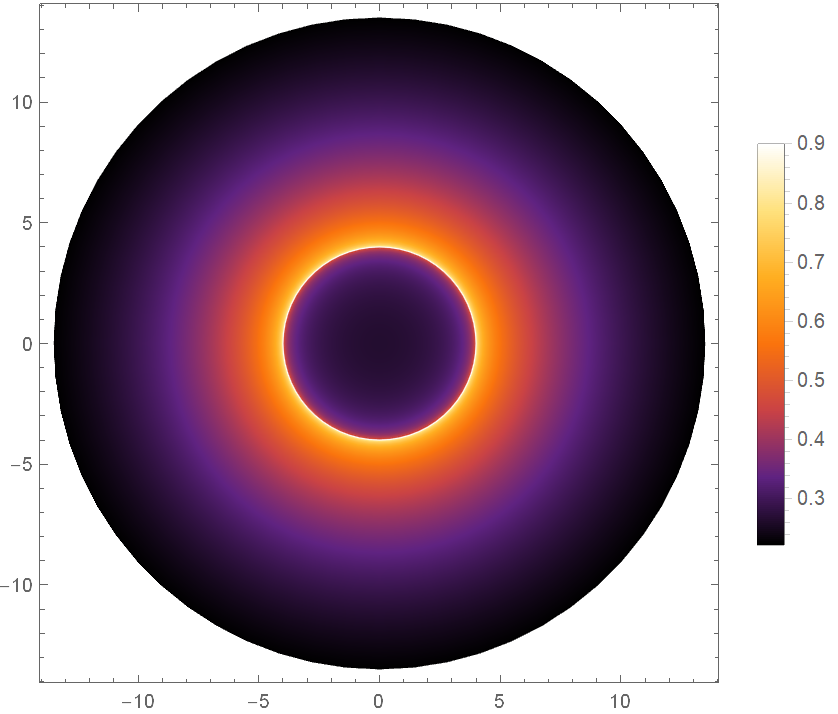}
    \caption{Profiles of the specific intensity $I({\tilde b})$ seen by a distant observer (Left panel) and its density function (Right panel) of the static traversable wormhole with parameter set: $\alpha = 1, \beta = 1, \Phi_0 = 1$ and $r_0 = 1$, and the shape function $b_{1}(r)$.}
    \label{fig_static_intensity1}
\end{figure}

\begin{figure}[h]
    \centering
    \includegraphics[width = 7 cm]{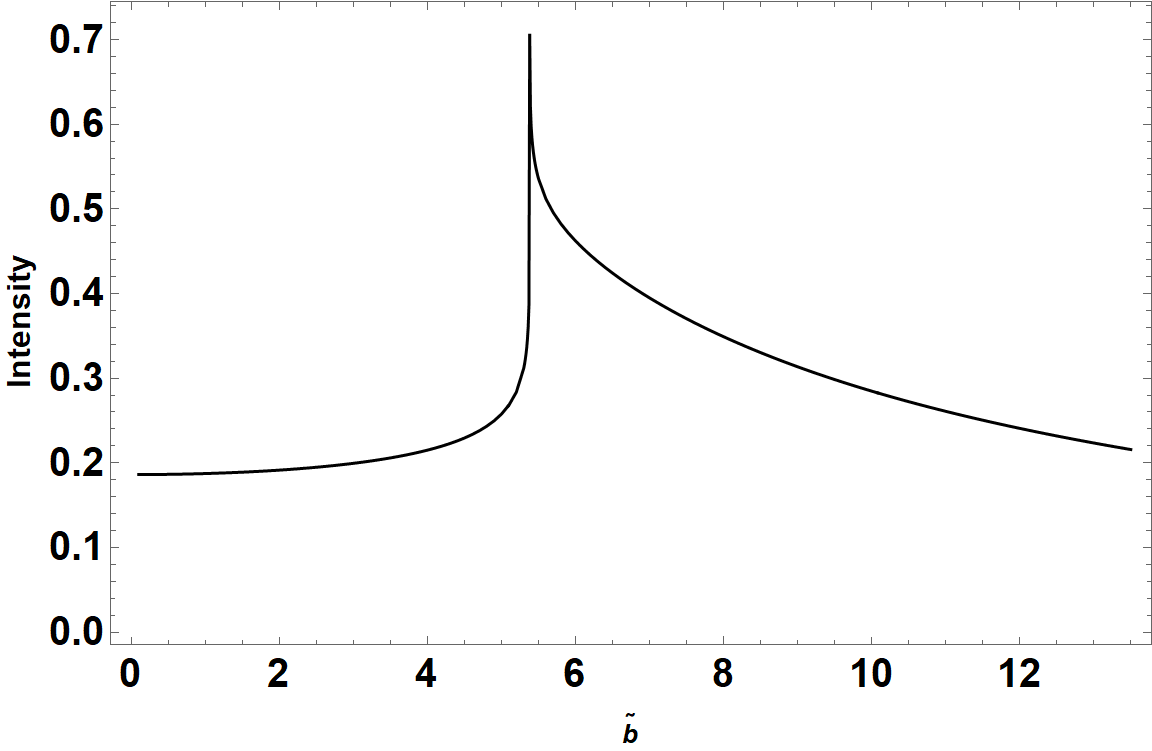}
    \includegraphics[width = 7 cm]{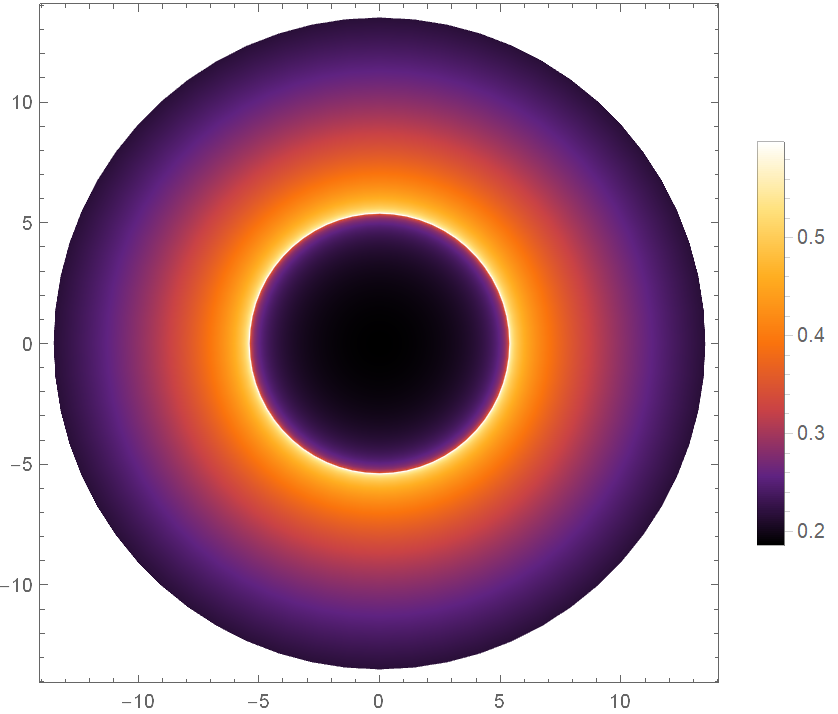}
    \caption{Profiles of the specific intensity $I({\tilde b})$ seen by a distant observer (Left panel) and its density function (Right panel) of the static traversable wormhole with parameter set: $\alpha = 1.5, \gamma = 1, \Phi_0 = 1$ and $r_0 = 1$, and the shape function $b_{2}(r)$.}
    \label{fig_static_intensity2}
\end{figure}

We perform numerical computations to determine the intensity of each ray's trajectory, utilizing the red-shift function detailed in Eq.~(\ref{redshift_function}), in combination with two distinct shape functions as outlined in Eq.~(\ref{shape_functions}). These computations employ parameter sets identical to those used in the previous subsections. From the left panel of Fig.~\ref{fig_static_intensity1} \& Fig.~\ref{fig_static_intensity2}, we observe that the intensity increases rapidly and then reaches a peak at $\Tilde{b}_{\rm ph}$, and then drops to lower values with increasing $\Tilde{b}$. This result is consistent with the right panel of Fig.~\ref{fig_static_intensity1} \& Fig.~\ref{fig_static_intensity2}. Since For $\Tilde{b} < \Tilde{b}_{\rm ph}$, most of the intensity originating from the accretion is absorbed by the wormholes. 

At $\Tilde{b} = \Tilde{b}_{\rm ph}$, the light ray undergoes multiple revolutions around the wormholes, resulting in the highest observable intensity. Conversely, when $\Tilde{b}> \Tilde{b}_{\rm ph}$, only refracted light contributes to the observer's intensity, and as the parameter $b$ increases, the contribution of refracted light diminishes. As the value of "$b$" increases, the refracted light becomes less. Consequently, when "$b$" reaches a sufficiently large value, the observed intensity completely disappears. In principle, the peak intensity at $\Tilde{b} = \Tilde{b}_{\rm ph}$ should be infinite since the light ray undergoes an infinite number of revolutions around the wormholes, accumulating an arbitrarily large intensity.

\subsection{Rotating case}

This case can be thought as a modification of the static one. It is straightforward to write a function $g$ as $g = \nu_0 / \nu_e = \left( e^{2 \Phi} - r^2 K^2 \omega^2 \right)^{1/2}$\,. Again setting $\omega=0$ and $K=1$, it reduces to the previous case. In this rotating case, we assume the same form of the emissivity $j(\nu_e) \propto \frac{\delta(\nu_e - \nu_r)}{r^2}$. The proper length measured in the rest frame of the emitter takes the form
\begin{eqnarray}
    dl_{\text{prop}} = \left( \frac{1}{1 - \frac{b}{r}} + r^2 K^2 \left( \left(\frac{d \phi}{dr} \right)^2 - 2 \omega \frac{d\phi}{dr} \frac{dt}{dr}  \right) \right)^{1/2}dr.
\end{eqnarray}
Notice that it is just a modification of the static case. Here $d\phi/dr$ is given by Eq.~(\ref{dpr}). The intensity in thei case can be written as
\begin{eqnarray}\label{in2}
    I(\nu_0) = - \int d \phi u'(\phi) \left( e^{2 \Phi} - \frac{K^2 \omega^2}{u^2} \right)^{3/2} \left( \frac{1}{1 - b u}  + K^2 u^2 \left( 1 - \frac{2 \omega}{\omega + \frac{u^2 e^{2\Phi}}{K^2 \left( E - \omega \right)}}  \right) \right)^{1/2}\,.
\end{eqnarray}
We will employ Eq.~(\ref{in2}) to investigate the shadow of the rotating wormholes. It is simple to verify that the expression given in Eq.~(\ref{in2}) reduces to that of the static case when setting $\omega=0$ and $K=1$.

\subsubsection{Ray tracing}

\begin{figure}
    \centering
    \includegraphics[width = 7 cm]{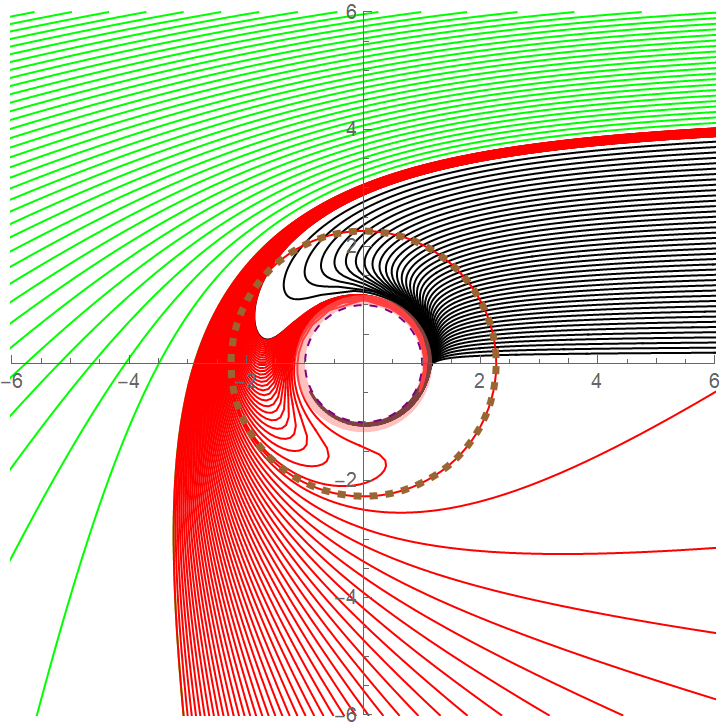}
    \includegraphics[width = 7 cm]{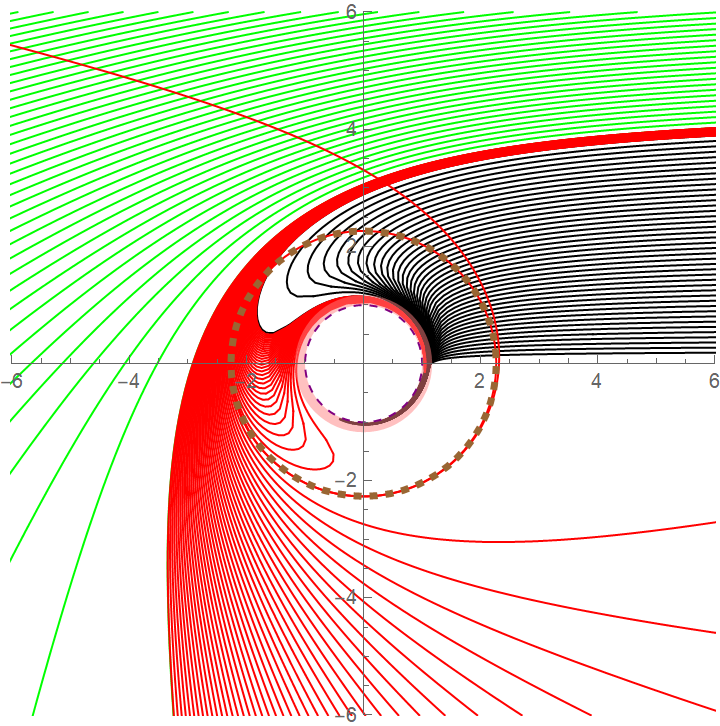}
    \caption{Figures show the ray tracing of the light geodesic around a rotating traversable wormhole with parameter set: $K = 1, a = -0.1, E = 1.5, \alpha = 1, \beta = 1, \Phi_0 = 1$ and $r_0 = 1$, and the shape function $b_{1}(1)$ (Left panel) and $K = 1, a = -0.1, E = 1.5, \alpha = 1.5$,\, $\gamma = 1, \Phi_0 = 1$ and $r_0 = 1$, and the shape function $b_{2}(r)$ (Right panel). The photon ring is symbolized by the dashed circle in brown, while the purple dashed circle indicates the wormhole's throat.}
    \label{fig_rotating_clockwise_raytracing1}
\end{figure}

In rotating black holes, it has been widely explored that geodesic equations can be numerically solved to illustrate ray tracing. Here, we use the same methodology when considering wormholes. We demonstrate the trajectory of light rays traveling around the rotating traversable wormholes, as shown in Fig.~\ref{fig_rotating_clockwise_raytracing1}. We consider again into two cases with different shape functions of wormholes introduced in Eq.~(\ref{shape_functions}). The black, red, and green lines correspond to $\Tilde{b} < \Tilde{b}_{\rm ph}, \Tilde{b} = \Tilde{b}_{\rm ph}$, and $\Tilde{b} > \Tilde{b}_{\rm ph}$, respectively. Using the center of the rotating wormholes as a reference point, one discover that the incoming light rays move in the anti-clockwise direction. For the case where $\Tilde{b} < \Tilde{b}{\rm ph}$, the light ray follows a trajectory into the wormhole, represented by the black curves in Fig.~\ref{fig_rotating_clockwise_raytracing1}. When $\Tilde{b}$ is exceptionally small, its trajectory is compelled to follow a clockwise orbit before eventually entering the wormhole's throat ($r_0 = 1$). Remarkably, when $\Tilde{b} \in (0, \Tilde{b}_{\rm ph})$, the light rays undergo a transition from counterclockwise to clockwise direction as they approach and enter the wormhole's throat. 

When $\Tilde{b} = \Tilde{b}_{\rm ph}$, the light ray is trapped in the photon ring presented in dashed brown circle or $\Tilde{b}_{\rm ph} = 4.36$ for both parameter sets. This rotating traversable wormhole has an ergosphere with an equatorial radius of $1.17$ ($\theta = \pi/2$). For the case of $\Tilde{b} > \Tilde{b}_{\rm ph}$, the light ray near the wormholes is refracted, which corresponds to the green rays in Fig.~\ref{fig_rotating_clockwise_raytracing1}. Similarly to the static case, when $\Tilde{b} > \Tilde{b}_{\rm ph}$, in order to plot the geodesic (the path of a light ray), we find a turning point, where the light ray changes its radial direction. The turning point is determined by ${\cal G}(u)=0$, where ${\cal G}(u)$ has been defined in Eq.~(\ref{metr}). The brown dashed circle represents the photon ring and the purple dashed circle represents the throat of the wormhole.

\subsubsection{Intensity}
\begin{figure}[h]
    \centering
    \includegraphics[width = 7 cm]{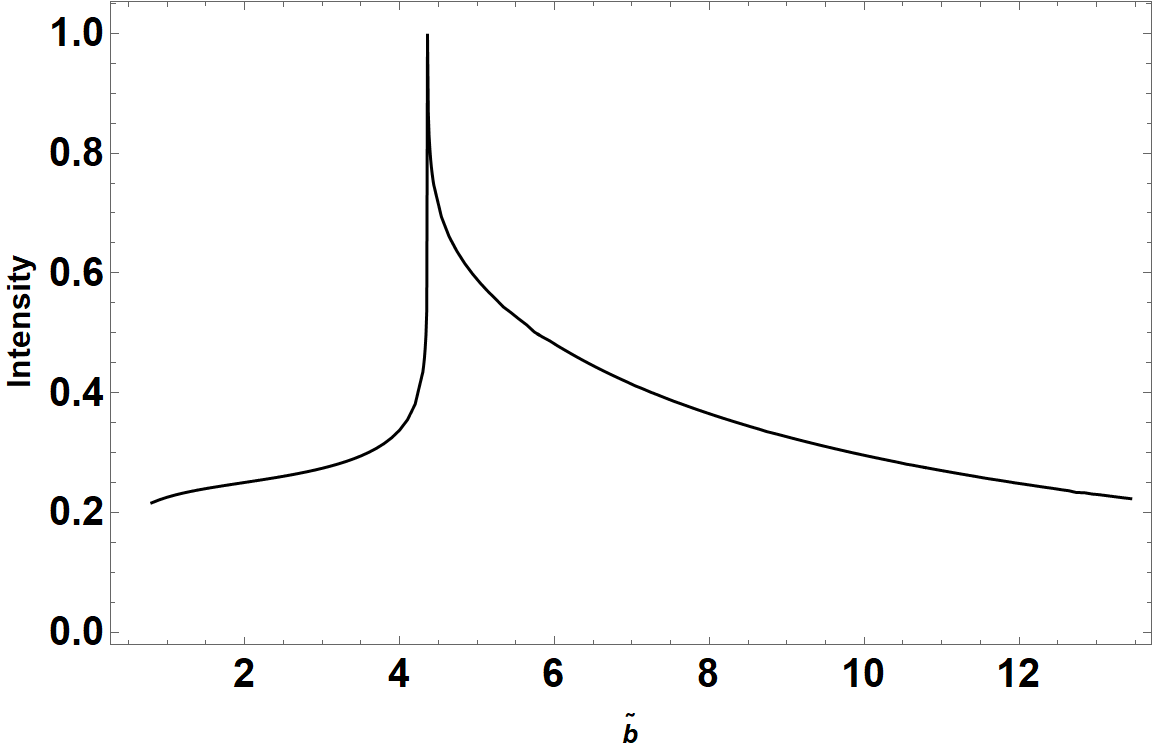}
    \includegraphics[width = 7 cm]{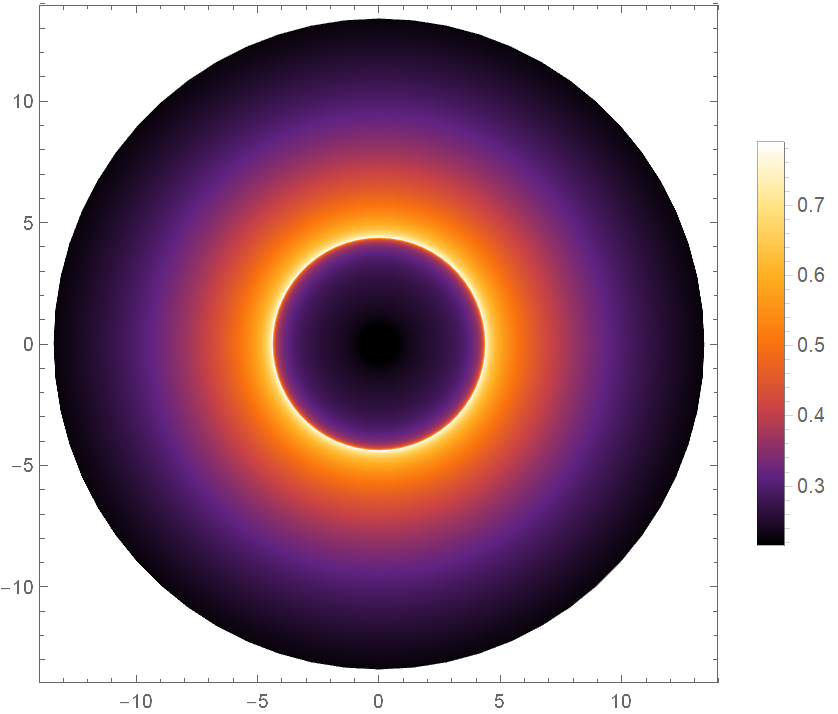}
    \caption{Profiles of the specific intensity $I({\tilde b})$ seen by a distant observer (Left panel) and its density function (Right panel) of the clock-wise rotating traversable wormhole with parameter set: $K = 1, a = -0.1, E = 1.5, \alpha = 1, \beta = 1, \Phi_0 = 1,$ and $r_0 = 1$, and the shape function $b_{1}(r)$.}
    \label{fig_rotating_clockwise_intensity1}
\end{figure}

\begin{figure}[h]
    \centering
    \includegraphics[width = 7 cm]{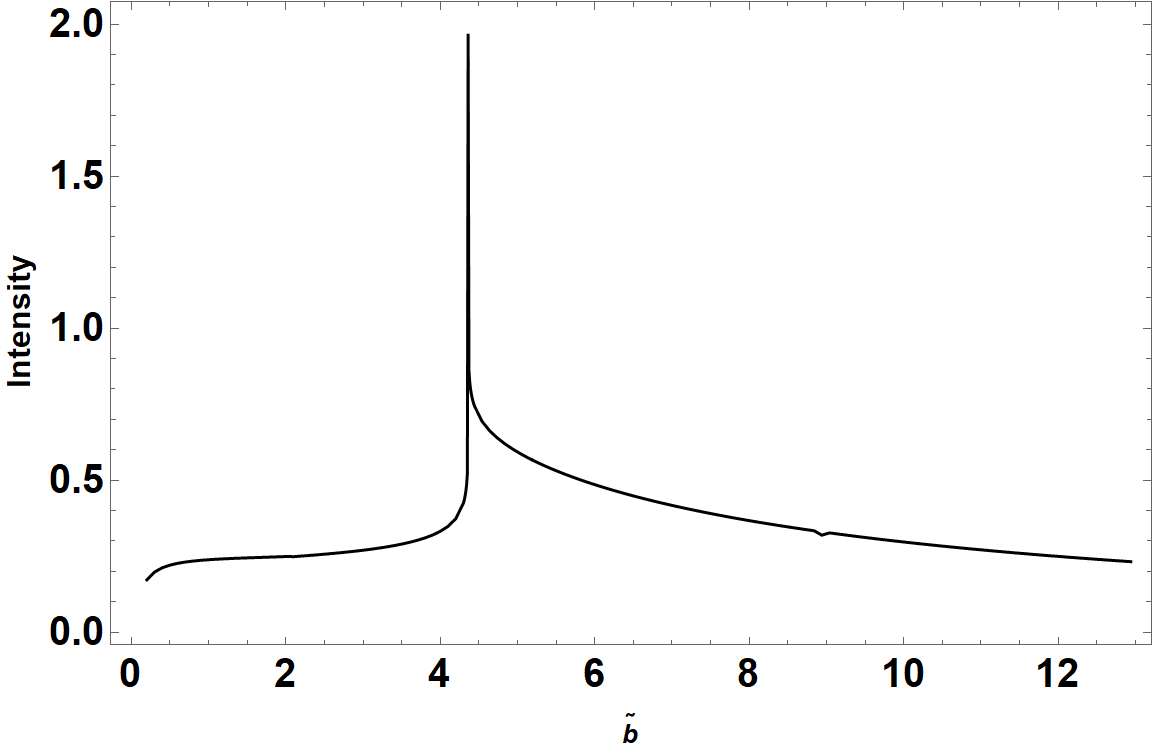}
    \includegraphics[width = 7 cm]{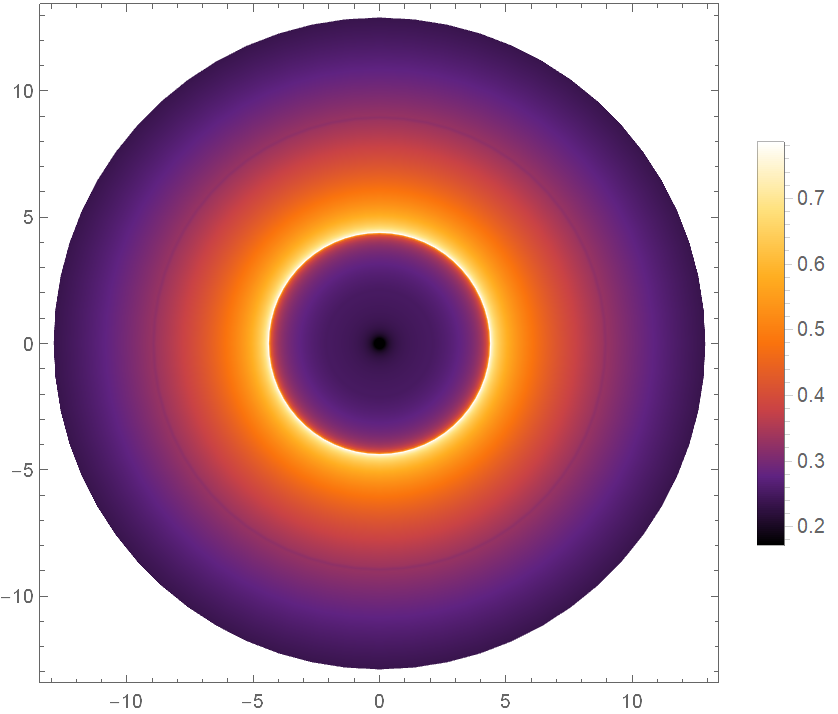}
    \caption{Profiles of the specific intensity $I({\tilde b})$ seen by a distant observer (Left panel) and its density function (Right panel) of the clock-wise rotating traversable wormhole with parameter set: $K = 1, a = -0.1$,\,$E = 1.5, \alpha = 1.5, \gamma = 1, \Phi_0 = 1$ and $r_0 = 1$, and the shape function $b_{2}(r)$.}
    \label{fig_rotating_clockwise_intensity2}
\end{figure}

In the case of rotation, we apply the same approach as described earlier to perform numerical calculations for determining the intensity of each ray's path. This involves using the red-shift function detailed in Eq.~(\ref{redshift_function}) in conjunction with two distinct shape functions as described in Eq.~(\ref{shape_functions}). The parameters used in these calculations are identical to those employed in previous sections. From the left panel of Fig.~\ref{fig_rotating_clockwise_intensity1} \& Fig.~\ref{fig_rotating_clockwise_intensity2}, it is evident that the intensity increases rapidly, reaching its peak at $\Tilde{b}_{\rm ph}$, and then decreases as $\Tilde{b}$ increases. This observation aligns with what we observed in the right panel of Fig.~\ref{fig_rotating_clockwise_intensity1} \& Fig.~\ref{fig_rotating_clockwise_intensity2}. For values of $\Tilde{b}$ less than $\Tilde{b}_{\rm ph}$, the intensity originating from accretion is mainly absorbed by the wormholes.

When $\Tilde{b} = \Tilde{b}_{\rm ph}$, the light ray embarks on numerous revolutions around the wormholes, resulting in the highest observable intensity. However, when $\Tilde{b}$ exceeds $\Tilde{b}_{\rm ph}$, only refracted light contributes to the observer's intensity, and with the parameter $b$ progressively increasing, the influence of refracted light diminishes correspondingly. As $b$ attains larger values, the contribution of refracted light dwindles significantly, eventually leading to the complete disappearance of observed intensity for sufficiently large $b$. In principle, the intensity peak at $\Tilde{b} = \Tilde{b}_{\rm ph}$ should be infinite because the light ray completes an infinite number of revolutions around the wormholes, accumulating an arbitrarily large intensity.

\section{Discussions and conclusions}\label{conc}
We have investigated the shadows and photon spheres for both static and rotating traversable wormholes. Specifically, we have considered a red-shift function, $\Phi(r)=\Phi_{0}\log \big(1-\frac{r_{0}}{r}\big)^{\alpha}$, in combination with two distinct shape functions, $b_{1}(r)=r\exp \big(-\beta (r-r_{0})\big)$ and $b_{2}(r)=r_{0} \left(\frac{r_{0}}{r}\right)^{\gamma}$. In so doing, we have first derived the null geodesic equation, allowing us to predict the light trajectory obtained by the Euler-Lagrange equation. In a static case, we have considered a usual static, spherically symmetric and asymptotically flat space-time. Having used the null geodesic equation, we could predict the light trajectory which illustrates the shadow and photon sphere of the wormholes. We have found that a critical value of the impact parameter of the case \#1, $\Tilde{b}_{ph, 1}=4.0$, which is less than that of the case \#2, $\Tilde{b}_{ph, 2}=5.3$.

In both static and rotating wormhole geometries, we have performed numerical computations to figure out the intensity of each ray's trajectory. Our results showed that at $\Tilde{b} = \Tilde{b}_{\rm ph}$, the trajectory of the light ray completes multiple revolutions around the wormholes, resulting in the highest observable intensity. In contrast, when $\Tilde{b}$ exceeds $\Tilde{b}_{\rm ph}$, only refracted light contributes to the observer's intensity, and as the value of the parameter '$b$' grows larger, the influence of refracted light diminishes. As the value of $b$ grows larger, the contribution from refracted light diminishes accordingly. Consequently, the observed intensity approaches zero for sufficiently large values of $b$. In principle, the peak intensity at $\Tilde{b} = \Tilde{b}_{\rm ph}$ should tend toward infinity because the light ray undergoes an infinite number of revolutions around the wormholes, progressively accumulating an arbitrarily large intensity.

Distinguishing between wormholes and black holes based on their shadows and photon sphere properties would be challenging, as both wormholes and black holes could produce similar observable effects in these aspects. Moreover, wormholes remain theoretical constructs, and their properties are not well-defined in the same way as black holes, which have been studied extensively in astrophysics. To distinguish between them conclusively, we would need more information and observations of these objects, should wormholes ever be confirmed to exist. Another interesting topic is that a study of particle trajectory is worth investigating. This may shed light on understanding the behavior of particles and gravitational effects near wormholes, see, for example, Ref.~\cite{Turimov:2022iff}.

It was noticed that the most recent report from the Sloan Digital Sky Survey Quasar Lens Search (SQLS) based on SDSS II marked a significant milestone by establishing the initial cosmological limitations on negative-mass compact objects and Ellis wormholes \cite{Takahashi:2013jqa}. However, ongoing or upcoming surveys, including Pan-Starrs \footnote{https://outerspace.stsci.edu/display/PANSTARRS/}, Dark Energy Survey \footnote{http://www.darkenergysurvey.org}, and the Large Synoptic Survey Telescope (LSST) \footnote{http://www.lsst.org/lsst}, are expected to identify a significantly greater number of gravitationally lensed quasars caused by foreground galaxies in comparison to the SDSS II. In line with our current work, we aim to further validate our findings by comparing them with the observed cosmic abundances of negative-mass compact objects. However, we leave these challenging issues for our future investigation.

\begin{acknowledgments}
T. Tangphati is financially supported by Research and Innovation Institute of Excellence, Walailak University, Thailand under a contract No. WU66267. The work of KB was partially supported by the JSPS KAKENHI Grant Number 21K03547.
\end{acknowledgments}

\end{document}